\documentclass[prl,showpacs,twocolumn,superscriptaddress,floatfix]{revtex4}
\usepackage{graphicx}
\usepackage{amsmath}
\usepackage{amsfonts}
\usepackage{amssymb}
\usepackage{bm}
\usepackage{psfrag}
\usepackage{epsfig}
\usepackage{amsmath}
\usepackage{sidecap}
\usepackage[figuresleft]{rotating}
\usepackage{caption}
\usepackage[]{graphicx}
\usepackage{lscape}

\begin{document}
\title {Transport through a molecular quantum dot in the polaron crossover regime}
\author{A. Zazunov}
\affiliation{LPMMC CNRS, 25 Avenue des Martyrs, 38042 Grenoble, France}
\affiliation{Centre de Physique Th\'eorique, Case 907, Luminy, 13288
Marseille Cedex 9, France}
\author{T. Martin}
\affiliation{Centre de Physique Th\'eorique, Case 907, Luminy, 13288
Marseille Cedex 9, France} \affiliation{Universit\'e de la
M\'edit\'erann\'ee, 13288 Marseille cedex 9, France}

\date{\today}
 \begin{abstract}
We consider resonant transport through a molecular quantum dot coupled to a local vibration mode.
Applying the non-equilibrium Green function technique in the polaron representation,
we develop a non-perturbative scheme to calculate 
the electron spectral function of the molecule
in the regime of intermediate electron-phonon coupling.
With increasing tunneling coupling to the leads, 
correlations between polaron clouds become more important
at relatively high temperature leading to a strong sharpening of the peak structure 
in the spectral function. The detection of such features in the current-voltage
characteristics is briefly discussed.
\end{abstract}

\maketitle

The possibility of using the degrees of freedom of molecules
contacted to metallic leads for nanoelectronics has been pointed out
in the last decade \cite{chen,park,bockrath,Joachim}. Molecular
devices typically consist of single molecules, carbon nanotubes,...,
connected to leads, where vibrational degrees of freedom and
electronic interactions operate. Their characterization via
transport remains an experimental and a theoretical challenge.
Depending on the strength of the electron-phonon coupling, various
approximation schemes have been developed for weak coupling
\cite{aleiner,mora,di_ventra} and for the intermediate to strong coupling,
where the electrons of the molecular quantum dot are
dressed by a polaron cloud \cite{shekhter,wingreen,schoeller,lundin,balatsky,alexandrov,flensberg,ratner,%
zazunov_ndc,zazunov_squeezing}.
The purpose of the present paper is to go beyond the existing polaron approaches to
molecular transport, allowing a better description of the
intermediate electron-phonon coupling regime.
While conventional perturbation theory is typically formulated with the help
of (single) phonon Green functions, for the intermediate coupling regime a more natural
choice is to introduce a Green function which describes collective phonon excitations
associated with the polaron cloud.
We show that the electron spectral function of the
molecular quantum dot can be evaluated systematically by diagram dressing of the polaron Green functions.
In the polaron crossover regime, where the coupling to the leads becomes comparable to
the phonon frequency and/or the strength of the electron-phonon coupling,
the phonon sidebands in the spectral function are found to remain sharp even at high temperature.
This implies that at intermediate coupling,
phonon features for the current through the molecule could
be more pronounced than expected. Pauli blocking of linewidths
in the strong coupling regime and at low temperature has recently been discussed \cite{flensberg}.

We consider a 
generic model which couples 
a single-electron level
to a local vibration mode with frequency $\Omega$.
The Hamiltonian of the molecule coupled via tunneling to two
metallic leads is given by $H = H_m + H_{leads} + H_T$,
with ($\hbar = 1$):
\begin{equation}
H_m = \left[ \, \epsilon_0 - g \left(b + b^\dagger \right) \right] d^\dagger d + \Omega \, b^\dagger b \, ,
\end{equation}
where $d$ ($b$) are annihilation operators for the dot electron (phonon),
$\epsilon_0$ is the bare dot level, and $g$ is the electron-phonon coupling.
The lead Hamiltonian is $H_{leads} = \sum_{j k} \xi_k \, c_{j k}^\dagger c_{j k}$,
where $c_{jk}$ is the electron operator for lead $j=l,r$. 
The tunneling Hamiltonian is $H_T = \sum_{j k} {\cal T}_{jk}^\ast \, c_{j k}^\dagger \, d + h.c.$,
where ${\cal T}_{jk}$ is the tunneling amplitude between lead $j$ and the molecule.
For simplicity, we ignore the spin degree of freedom assuming
a resonant tunneling situation in the Coulomb blockade regime, far from the Kondo regime.
The bias voltage across the junction, $V = V_l - V_r$, is imposed by shifting the chemical potentials
of the leads, $\mu_j = e V_j$. Throughout the paper, we consider the symmetric case,
${\cal T}_{jk} = {\cal T}_k$, assuming the electron-hole symmetry to be held in the presence of phonons (see below).
In the steady state, the current through the molecule can be expressed as \cite{MW}:
\begin{equation}
I(V) = e \int {d \omega \over 4 \pi } \, \left[ f_l(\omega ) - f_r
(\omega ) \right] \, \Gamma(\omega ) \, A(\omega ) \,,
\label{current}
\end{equation}
where $A(\omega )$ is the Fourier transform of the electron spectral function of the dot,
$A(t - t') = -i \langle \{ d(t) ,\, d^\dagger(t') \} \rangle$,
$f_{l/r}(\omega) = f(\omega \mp e V/2)$ are Fermi distributions in the leads,
$f(\omega) = \left( e^{\beta \omega} + 1 \right)^{-1}$, and
$\Gamma(\omega ) = 2 \pi \sum_k |{\cal T}_k|^2 \, \delta \left( \omega -  \xi_k \right)$
is the tunneling half-width.

We eliminate the electron-phonon coupling term in $H_m$ by using the polaron unitary transformation $U$ \cite{mahan}:
$\tilde{H} = U^\dagger H U  = \tilde{H}_m + H_{leads} + \tilde{H}_T$
with $U = e^{- i \alpha p \, d^\dagger d }$,
$p = i \left( b^\dagger - b \right)$, and $\alpha = g / \Omega$.
The diagonalized Hamiltonian of the molecule reads:
$\tilde{H}_m = \varepsilon \, d^\dagger d + \Omega \, b^\dagger b$,
where $\epsilon = \epsilon_0 - g^2/ \Omega$ is the renormalized level energy.
The assumed above electron-hole symmetry implies that $\epsilon = 0$
which can be achieved by applying a gate voltage.
The transformed tunnel Hamiltonian becomes dressed by the vibrations,
$\tilde{H}_T = \sum_{j k} {\cal T}_k^\ast \, c_{j k}^\dagger D + h.c.$,
where $D=dX$ and $X =  e^{- i \alpha p}$ is the polaron cloud operator.
This form of $\tilde{H}_T$ suggests an effective phonon-mediated coupling between the dot and lead electrons.
In the polaron representation, the spectral function reads
$A(t - t') = -i \langle \{ D(t) ,\, D^\dagger(t') \} \rangle$.

We thus introduce the Keldysh Green function for the polaron,
${\cal G}^{s s'}(t - t') = - i \, {\langle \, T_K \{ \, D^{s}(t) D^{\dagger s'}(t')  \} \rangle}$,
where $s, s' = \left\{1,2\right\}$ stand for the forward (1) and backward (2)
branches of the Keldysh contour, and $T_K$ is the Keldysh time-ordering.
Averaging over the leads results in:
\begin{equation}
{\cal G}^{s s'}(t - t') = - i \,
{\langle \, T_K \{ S_T \, D^{s}(t) D^{\dagger s'}(t')  \} \rangle}_0 \,,
\label{calG}
\end{equation}
where ${\langle \, ...  \rangle}_0 =
$
is a trace over the states of the molecule only,
and the time-dependence of $D(t)$ is given in the interaction picture with respect to $\tilde{H}_m$.
In Eq. (\ref{calG}), the effective S--matrix operator reads:
\begin{equation}
S_T = T_K \exp \left\{- i \int_{-\infty }^{+\infty } d \tau d \tau' \, W(\tau, \tau') \right\} \,,
\label{Smatrix}
\end{equation}
\begin{equation}
W(\tau, \tau') = \sum_{s s'} D^{\dagger s}(\tau ) \Sigma_0^{s s'}(\tau - \tau') D^{s'}(\tau') \,,
\end{equation}
where the electron tunneling self-energy $\Sigma_0$ is
built up from the Keldysh Green functions for the uncoupled leads,
$g^{s s'}_{jk}(\tau)$, $\Sigma_0^{s s'}(\tau) = 
(-1)^{s+s'} \sum_{j k} | {\cal T}_k |^2 \, g^{s s'}_{jk}(\tau)$,
and can be expressed in terms of $\Gamma (\omega)$ and $f_j(\omega)$ by virtue of
$\Sigma_0^{<,>}(\omega) = \pm i \Gamma(\omega) \sum_j \, f_j( \pm \omega)$.

We now proceed with the perturbation theory in the lead self-energy.
The main challenge resides in the fact that because of the multi-phonon operator factor $X$
accompanying the $d$ operator in $S_T$,
the Wick theorem is not applicable to express ${\cal G}$ in terms of the electron and phonon self-energies.
It is convenient to introduce a Keldysh Green function describing the polaron cloud:
\begin{equation}
\Lambda^{s s'}(t)= {\langle \, T_K \left( X^s(t) X^{\dagger s'}(t') \right)  \rangle}_0
= \exp\{\, - \alpha^2 \Phi^{s s'}(t) \} \;,
\end{equation}
with $\Phi^{s s'}(t)= (1/2) {\langle \, T_K \left( p^s(t) - p^{s'}(0) \right)^2 \rangle}_0$.
A typical approximation \cite{kuochang,lundin,alexandrov,balatsky},
which is called  the ``single particle approximation'' (SPA) in Ref. \onlinecite{flensberg},
consists in factorizing the electron Green function as
\begin{equation}
{\cal G}^{s s'}(t-t') \simeq {\bf G}^{s s'}(t - t') \, \Lambda^{s s'}(t - t') \;,
\label{calG_Lundin}
\end{equation}
where ${\bf G}$ obeys the Dyson equation (in the Fourier representation)
${\bf G}(\omega ) = G(\omega ) +
G(\omega ) \, \Sigma_0(\omega ) \, {\bf G}(\omega )$,
with $G$ the bare Green function of the isolated dot (in the absence of phonons).
This decoupling of electron and phonon dynamics can be justified in the anti-adiabatic regime \cite{feinberg},
where $\Omega$ (and/or $g$) is large compared to the dot tunneling rates.
In this regime, the phonon mode is fast and reacts instantaneously (on the time scale of $\Gamma^{-1}$ )
to the presence/absence of an electron on the molecule.
Eq. (\ref{calG_Lundin}) leads to the following expression for the spectral function:
\begin{equation}
A_{SPA}(t) = {\bf G}^{>}(t) \, \Lambda^>(t) - {\bf G}^{<}(t)\, \Lambda^<(t) \,,
\label{Aspa}
\end{equation}
where $\Lambda^>(t) = \Lambda^<(-t) = e^{-\alpha^2 \Phi^>(t)}$
with
\begin{equation}
\Phi^>(t)= (2 N + 1) (1 - \cos \Omega \tau) + i \sin \Omega \tau \,,
\label{Phi}
\end{equation}
and $N = \left( e^{\beta \, \Omega } - 1 \right)^{-1}$.
\begin{figure}[h!]
\scalebox{0.3}{\includegraphics{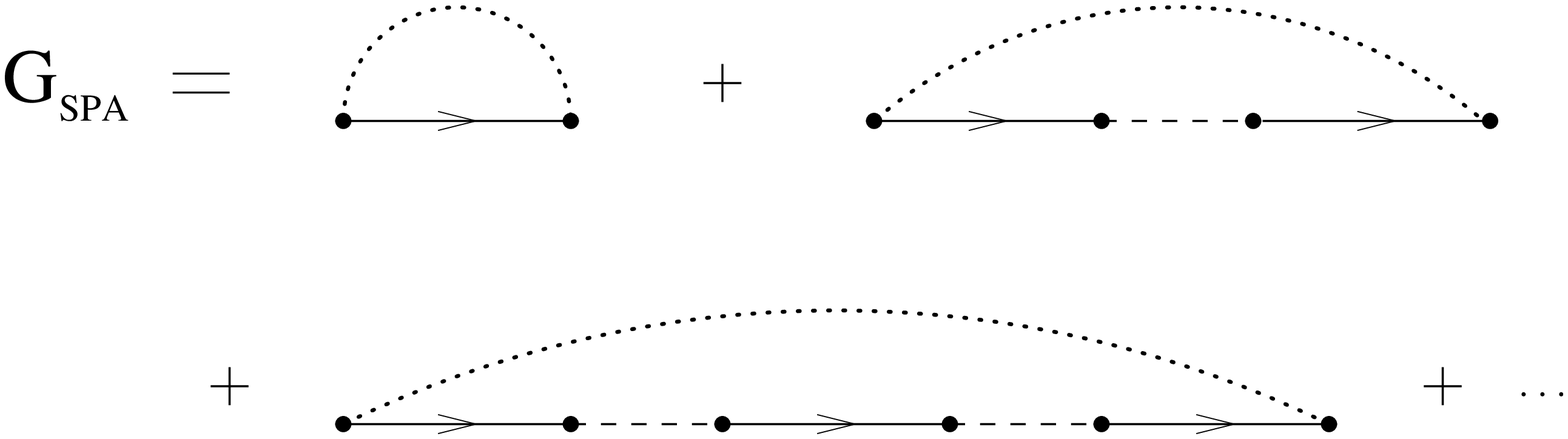}}
\caption{ The Dyson series for ${\cal G}$ 
corresponding to the single-particle approximation, Eq. (\ref{calG_Lundin}).
Solid lines represent the bare electron propagator $G$, dashed lines the self-energy $\Sigma_0$ (lead electrons), and dotted lines indicate the phonon-cloud dressing propagator $\Lambda$.}
\label{figDiagram}
\end{figure}
The Feynman diagrams associated with the SPA
are depicted in Fig. \ref{figDiagram}; note that in each order the phonon cloud
connects only the end points of the full fermionic lines.
In \cite{flensberg,ratner}, the equation of motion method was used to go beyond this approximation.
Below, we compare the SPA with our own scheme,
labeled as the  ``nearest-neighbor crossing approximation'' (NNCA),
which incorporates the effects of the polaron cloud in the vertex part of the Green function.

First, to zeroth order in the tunneling rates,
the Keldysh Green function for the polaron is given by an expression similar to Eq. (\ref{calG_Lundin}),
\begin{equation}
{\cal G}_0^{s s'}(t-t') =  G^{s s'}(t-t') \, \Lambda^{s s'}(t-t'),
\end{equation}
but with the bare propagator $G$. In this so-called atomic limit,
${\cal G}_0$ describes the localized electron state dressed by vibrations.
This term corresponds to the first diagram in the series shown in Fig. \ref{figDiagram}.
Next, to first order in the tunneling rate, one obtains:
${\cal G}_1^{s s'}(t-t') = \tilde{{\cal G}}_1^{s s'}(t - t') \, \Lambda^{ss'}(t - t')$,
with
\begin{eqnarray}
\tilde{{\cal G}}_1^{s s'}(t - t') =
\int d \tau_1 d \tau_2 \sum_{s_1 s_2} \,
{\cal G}_0^{s s_1}(t - \tau_1) \, \left[ \Lambda^{s s_2}(t-\tau_2) \right]^{-1}
\hspace{-0.9 cm} \nonumber \\
\times \, \Sigma^{s_1 s_2}(\tau_1 - \tau_2) \, \left[ \Lambda^{s_1 s'}(\tau_1 - t') \right]^{-1}
\, {\cal G}_0^{s_2 s'}(\tau_2 - t') \;,
\label{tildecalG1}
\end{eqnarray}
where we have introduced the dressed self-energy
$\Sigma^{s_1 s_2}(\tau) = \Sigma_0^{s_1 s_2}(\tau) \, \Lambda^{s_1 s_2}(\tau)$.
We note that in Eq. (\ref{tildecalG1}), the phonon-cloud propagator $\Lambda^{-1}$
plays the role of a ``phonon undressing''  factor.

Although Eq. (\ref{tildecalG1}) looks rather complicated,
it can be drastically simplified
if one assumes that ${\rm Re} \, \Lambda \gg  {\rm Im} \, \Lambda$,
which is equivalent to saying that $\Lambda^K \gg \Lambda^R - \Lambda^A$,
where $\Lambda^{R/A/K}$ is the retarded/advanced/Keldysh component of $\Lambda$.
Neglecting ${\rm Im} \, \Lambda$ turns out to be
a good approximation for the case of intermediate electron-phonon coupling,
$\alpha < 1$.
\begin{figure}[h!]
\scalebox{0.25}{\includegraphics{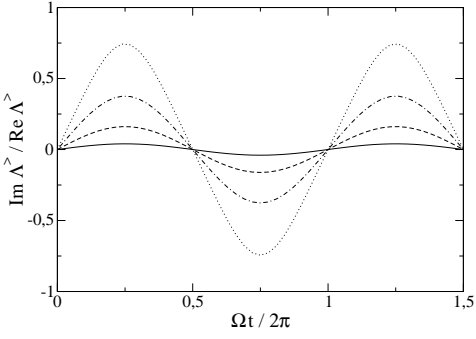}}
\caption{Ratio between the imaginary and real parts of $\Lambda^>$ as a function of time $t$
for $\alpha=0.8$ (dotted line), $0.6$ (dashed dotted line), $0.4$ (dashed line) and 
$0.2$ (full line).
\label{figImRe}
}
\end{figure}
This is illustrated in Fig. \ref{figImRe}, where it is shown that the real part of $\Lambda$
dominates with decreasing electron-phonon coupling $\alpha$.
As seen from Eq. (\ref{Phi}), this result does not depend on temperature.
Noticing that
\begin{equation}
\Lambda^{s s'} = (1/2) \Lambda^K +
(1/2) \left( \Lambda^{s s'} - \Lambda^{\bar{s} \bar{s}'} \right) \;,
\end{equation}
with $\bar{s} = 2(1)$ for $s = 1(2)$, and neglecting the last term
in the above expression  which is proportional to ${\rm Im} \, \Lambda$, 
one finds from Eq. (\ref{tildecalG1}) for the retarded component:
\begin{eqnarray}
\tilde{{\cal G}}_1^R(t - t') =
\int d \tau_1 d \tau_2 \,
{\cal G}_0^R(t - \tau_1) \, \lambda^{-1}(t-\tau_2)
\hspace{0.5cm} \nonumber\\
\times \; \Sigma^R(\tau_1 - \tau_2) \, \lambda^{-1}(\tau_1 - t') \,
{\cal G}_0^R(\tau_2 - t') \;,
\label{tildecalGR1}
\end{eqnarray}
where for notation convenience we have introduced $\lambda (\tau) = \Lambda^K(\tau) /2$.

\begin{figure}[h!]
\scalebox{0.3}{\includegraphics{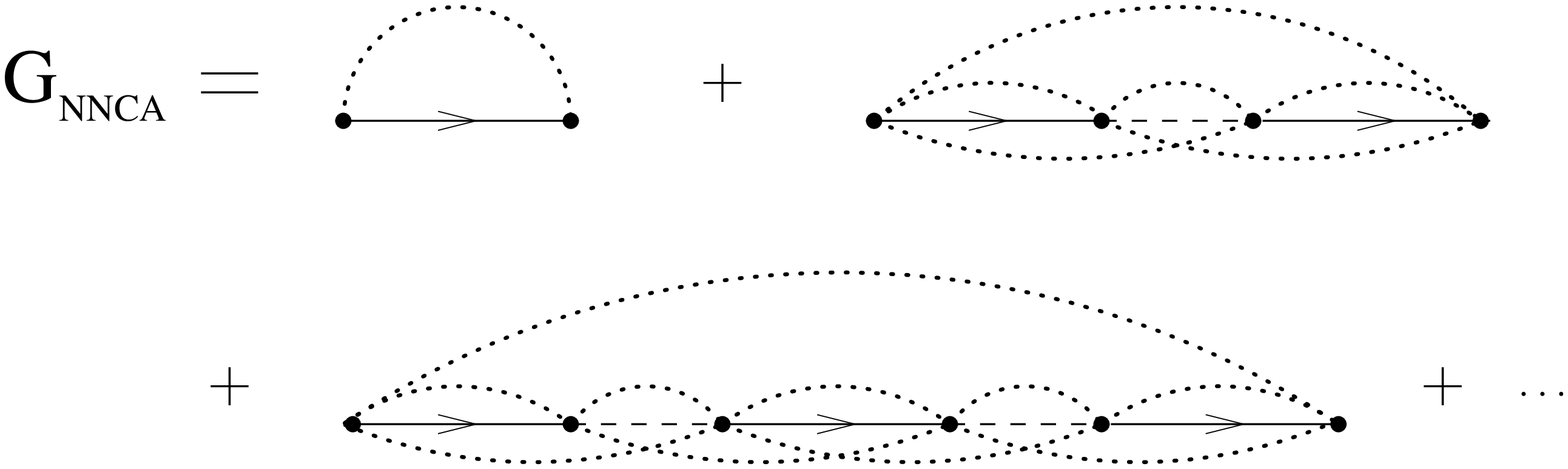}}
\caption{ \label{figNNCA}
Diagrammatic representation of the NNCA.
As in Fig. \ref{figDiagram}, dotted lines above the fermionic line represent
the phonon-cloud dressing propagator $\Lambda$, while 
dotted lines crossing below the fermionic line represent 
the undressing corrections described by $\lambda^{-1}$.}
\end{figure}

Now, going to higher order terms ($n>1$) of perturbation theory in the lead coupling,
we replicate the structure of the first order diagram ($n = 1$)
keeping only the nearest-neighbor crossing lines ($\lambda^{-1}$)
connecting the dressed electron propagators (${\cal G}_0$) and 
the dressed self-energies ($\Sigma$).
By doing this, we take into account correlations between phonon clouds
corresponding to two sequential tunneling of the lead electrons,
which is happening on the time scale of the order of $\Gamma^{-1}$.
Noticing that the general solution for the Green function can be written as
\begin{equation}
{\cal G}^R(t-t') = {\cal G}_0^R(t-t') + \lambda(t - t') \, \tilde{{\cal G}}^R(t - t') \;,
\label{NNCAcalGR}
\end{equation}
the $n$-order term for $\tilde{{\cal G}}^R$ in the NNCA reads ($n \geq 2$):
\begin{eqnarray}
\tilde{{\cal G}}_n^R(t - t') =
\int d \tau_1 d \tau_2 \,...\,d \tau_{2n} \,
{\cal G}_0^R(t - \tau_1) \,
\hspace{1cm}\nonumber\\
\times \; \lambda^{-1}(t-\tau_2) \, 
\Sigma^R(\tau_1 - \tau_2) \, \lambda^{-1}(\tau_1 - \tau_3) \,...
\hspace{1cm} \nonumber \\
... \; \Sigma^R(\tau_{2n-1} - \tau_{2n}) \,
\lambda^{-1}(\tau_{2n-1} - t') \, {\cal G}_0^R(\tau_{2n} - t') \,.
\label{tildecalGRn}
\end{eqnarray}
The Feynman diagrams associated with this approximation scheme are illustrated in Fig. \ref{figNNCA}.
Going to the Fourier representation and summing up all the terms of perturbation theory, one finds:
\begin{equation}
\tilde{{\cal G}}^R(\omega) = \tilde{{\cal G}}_0^R(\omega) \,
\Pi^R(\omega)
\, \tilde{{\cal G}}_0^R(\omega) \;,
\label{fullcalGR}
\end{equation}
where $\tilde{{\cal G}}_0^R(\omega)$ is the Fourier transform of
\begin{equation}
\tilde{{\cal G}}_0^R(\tau) = \lambda^{-1}(\tau) \, {\cal G}_0^R(\tau) \;,
\label{tildecalG0R}
\end{equation}
$\Pi^R$ satisfies the Dyson equation,
\begin{equation}
\Pi^R(\omega) = \Sigma^R(\omega) + \Sigma^R(\omega) \, K^R(\omega) \, \Pi^R(\omega) \;,
\label{PiRDysoneq}
\end{equation}
and $K^R(\omega)$ is the Fourier transform of
\begin{equation}
K^R(\tau)= \lambda^{-2}(\tau) \, {\cal G}_0^R(\tau) \;.
\label{KR}
\end{equation}
Eqs. (\ref{NNCAcalGR})--(\ref{KR}) constitute the basis of the NNCA
approximation developed here.

\begin{figure}[h!]
\scalebox{0.3}{\includegraphics{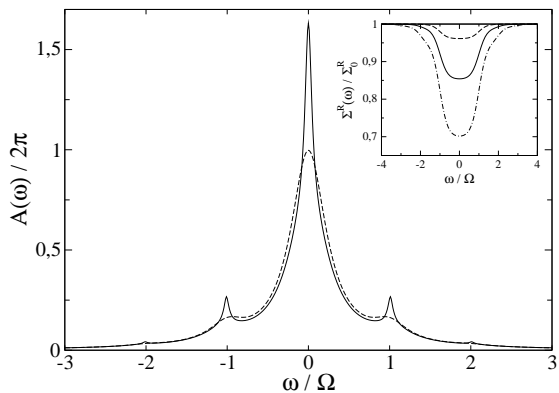}}
\caption{ \label{low}
The spectral function at $T/\Omega = 0.4$ 
calculated within the NNCA (solid line) and SPA (dashed line)
for $\alpha=0.4$, $2 \Gamma /\Omega = 0.5$, $\epsilon = 0$, and $V=0$.
Inset: Renormalized lead self-energy for (from top to bottom) $\alpha = 0.2$, $0.4$, and $0.6$.
The other parameters are the same.
}
\end{figure}

Although, after dressing, the retarded self-energy
$\Sigma^R(\tau) = \Theta(\tau) \left[ \Lambda^>(\tau) \Sigma_0^>(\tau) - 
\Lambda^<(\tau) \Sigma_0^<(\tau) \right]$ becomes voltage-dependent, 
inspection of this expression
in the wide-band limit, where the energy dependence of $\Gamma(\omega)$ is neglected,
shows that for the intermediate coupling $\alpha \lesssim 0.5$
the magnitude of $\Sigma^R$ is only slightly decreased compared to the bare one,
$\Sigma^R_0 =- i \Gamma$ (see inset of Fig. \ref{low}).
As a result, in the wide-band limit considered here, 
the voltage dependence of the spectral function 
can be discarded.

The difference between the NNCA and the SPA is hidden in the kernel $K^R$
entering the Dyson equation (\ref{PiRDysoneq}). To show this we first notice that
the expression for $\tilde{{\cal G}}_0^R$,
$\tilde{{\cal G}}_0^R(\tau) =
\lambda^{-1}(\tau) \Theta(\tau) \left[
G^{>}(\tau) \Lambda^{>}(\tau) - G^{<}(\tau) \Lambda^{<}(\tau) \right]$,
can be approximated by the bare Green function, $\tilde{{\cal G}}_0^R \approx G^R$,
which is valid up to the small term of the order of ${\rm Im} \Lambda / {\rm Re} \Lambda$.
As a result, the kernel $K^R$, Eq. (\ref{KR}), can be viewed as the ``anti-dressed''
propagator, $K^R(\tau) \approx \lambda^{-1}(\tau) \, G^R(\tau)$.
This expression is different by the factor $\lambda^{-1}$ from 
the corresponding expression in the SPA,
$K^R_{SPA} = G^R$.
This can be checked by replacing $K^R$ in Eq. (\ref{PiRDysoneq}) by $G^R$,
which results in the SPA expression for the retarded Green function:
\begin{equation}
{\cal G}^R_{SPA}(t-t') \approx \lambda(t - t') \, {\bf G}^R(t - t') \;,
\end{equation}
with ${\bf G}^R(\omega) = \left( \omega - \epsilon - \Sigma^R(\omega) \right)^{-1}$.

\begin{figure}[h!]
\scalebox{0.3}{\includegraphics{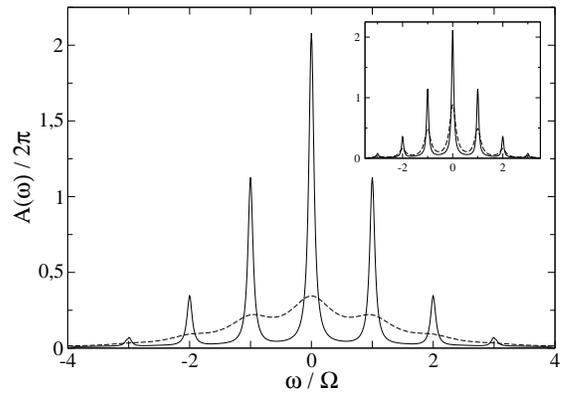}}
\caption{ \label{high}
Same as Fig. \ref{low} but for $T/\Omega = 4$ and $2 \Gamma /\Omega = 0.8$.
Inset: same as the main figure but for $2 \Gamma /\Omega = 0.2$.
}
\end{figure}

In Figs. \ref{low} and \ref{high} we plot the results for the spectral function
$A(\omega) = - 2 {\rm Im} \,{\cal G}^R(\omega)$ calculated within the NNCA
by solving numerically Eqs. (\ref{NNCAcalGR})--(\ref{KR}).
For comparison we also show the results corresponding to the SPA based on Eq. (\ref{Aspa}).
In all cases we take $\alpha = 0.4$ and $\epsilon = 0$.
In the low temperature regime (Fig. \ref{low}),
one sees that the NNCA predicts a slightly sharper
central peak and more pronounced phonon sidebands than expected in the SPA at relatively
high transparencies. In this regime, the peak sharpening can be attributed
to the suppression of the tunneling self-energy by the phonon cloud, which 
pins the dressed electron on the dot (inset of Fig. \ref{low}).
Because of the low temperature, only the single-phonon peaks are clearly visible.
With increasing temperature, the occupation probability of phonon sidebands increases,
and more satellite peaks appear in the spectral function (Fig. \ref{high}).
Although for small $\Gamma$, as expected,
the NNCA and SPA results are hardly distinguishable (see inset of Fig. \ref{high}),
the difference is clearly seen for the higher transparency (the main plot of Fig. \ref{high}).
According to the NNCA, all excited satellite peaks remain sharp and
are rather robust with respect to increasing $\Gamma$.
This is the regime where the SPA breaks down because the tunneling rate ($2 \Gamma$) becomes comparable to $\Omega$.
In this transition region, increasing the temperature acts in favor of
the electron-phonon interaction which tends to localize the dot electron,
leading to a well pronounced peak structure in the spectral function.
With increasing the $\Gamma$, such that $2 \Gamma > \Omega$,
we expect that the NNCA will not be sufficient, and the higher order correlations
between the polaron clouds should also be included.

Finally, we remark that the effect of suppression of the tunnel broadening in the spectral function
at high temperature can be observed directly in the measurement of the differential conductance
keeping the leads at very low temperature ($T \ll \Omega$), so that $dI/dV \propto A(eV)$,
but varying the local temperature of the molecule, for instance, by heating it with a laser.

We thank D. Feinberg for useful discussions.


\end{document}